\documentclass[aps,prl,preprint,superscriptaddress]{revtex4-2}
\usepackage{dsfont}
\usepackage{graphicx}% Include figure files
\usepackage{dcolumn}% Align table columns on decimal point
\usepackage{bm}% bold math
\usepackage[colorlinks=true,allcolors=blue]{hyperref}
\usepackage{color}
\usepackage{siunitx,tcolorbox,lastpage,amsmath,csquotes}

\begin{document}
	
	\title{Ultrafast signatures of Dirac / flat-band hybrid states from time-resolved ARPES}
	
	
	
	\author{Maria-Elisabeth Federl} \thanks{These authors contributed equally.}
	\author{Johannes Gradl} \thanks{These authors contributed equally.}	
	\author{Franziska Bergmeier}
	\author{Leonard Weigl}
  \affiliation{Department for Experimental and Applied Physics, University of Regensburg, 93040 Regensburg, Germany}
	
	\author{Lukas Bruckmeier}
	\affiliation{Department for Experimental and Applied Physics, University of Regensburg, 93040 Regensburg, Germany}
	\affiliation{Institut f\"ur Experimentelle und Angewandte Physik, Christian-Albrechts-Universität zu Kiel, 24098 Kiel, Germany}
	\affiliation{Ruprecht Haensel Laboratory, Deutsches Elektronen-Synchrotron DESY, 22607 Hamburg, Germany}
	
	\author{Theresa Glaser}
	\affiliation{Department for Experimental and Applied Physics, University of Regensburg, 93040 Regensburg, Germany}
	
	\author{Zamin Mamiyev}
	\author{Christoph Tegenkamp}
	\affiliation{Institute of Physics, Chemnitz University of Technology, 09126 Chemnitz, Germany}
	
	\author{Niclas Tilgner}
	\author{Thomas Seyller}
	\affiliation{Institute of Physics, Chemnitz University of Technology, 09126 Chemnitz, Germany}
	\affiliation{Research Center for Materials, Architectures and Integration of Nanomembranes (MAIN), 09126 Chemnitz, Germany}
	
	\author{Teresa Tschirner}
	\affiliation{Physikalisch-Technische Bundesanstalt, 38116 Braunschweig, Germany}
	
	\author{Domenica Convertino}
	\author{Stiven Forti}
	\author{Camilla Coletti}
	\affiliation{Center for Nanotechnology Innovation@NEST, Istituto Italiano di Tecnologia, 56127 Pisa, Italy}
	
	\author{Isabella Gierz}
	\email[]{isabella.gierz@ur.de}
	\affiliation{Department for Experimental and Applied Physics, University of Regensburg, 93040 Regensburg, Germany}
	
	
	\date{\today}
	
	\begin{abstract}
	
Hybridization of highly itinerant Dirac electrons with localized flat-band states is predicted to yield emergent phenomena such as exotic heavy-fermion behaviour. Epitaxial graphene on two-dimensional adsorbate structures on SiC(0001), which host flat bands, offers a promising platform to explore these effects. However, direct experimental evidence of interlayer hybridization in such systems has so far been lacking. Here, we address this gap using time- and angle-resolved photoemission spectroscopy (trARPES) where interlayer hybridization manifests in three key observations: (1) accelerated Dirac-carrier relaxation arising from additional electronic and phononic decay channels provided by the flat-band subsystem, (2) transient charging of the Dirac cone enabled by direct optical excitation from the flat bands, and (3) ultrafast back-transfer of charge into the flat bands on timescales governed by the interlayer coupling strength. We further demonstrate that the degree of hybridization can be tuned via the atomic number of the atoms intercalated at the graphene-SiC interface, establishing a controllable platform for investigating exotic correlated ground states.
	
	\end{abstract}
	
	\maketitle

\section{Introduction}

Dirac cones host relativistic, effectively massless charge carriers with high mobility \cite{wallace1947band, novoselov2004electric, berger2004ultrathin, novoselov2005two}, while flat bands are characterized by quenched kinetic energy and enhanced electron-electron interactions \cite{SunPRL2011, LeykamTaylorFrancis2018}. When these two contrasting regimes are brought into proximity, theory predicts the emergence of hybrid quasiparticles with unusual properties, including heavy-fermion-like behaviour \cite{song2022magic,cualuguaru2023twisted, herzog2025topological,witt2026real-space}.

Prominent examples of such physics have been explored in crystalline materials featuring Kagome lattices, such as the AV$_3$Sb$_2$ (A = K, Rb, Cs) family, which exhibits charge density waves and superconductivity \cite{OrtizPRM2019, OrtizPRL2020}, as well as Kagome magnets, where flat bands and magnetic frustration have been linked to spin-liquid behaviour and nontrivial topology \cite{BalentsNature2010, LinPRL2018, YeNature2018}. Related effects have also been demonstrated in artificial Lieb lattices, realized in photonic and ultracold-atom architectures \cite{LiebPRL1989, Guzman-SilvaIOP2014, MukherjeePRL2015, TaieSciAdv2015}, as well as in moiré systems such as magic-angle twisted bilayer graphene, where the Dirac dispersion collapses into narrow bands due to the moir\'e potential \cite{CaoNature2018, CaoNature2018_supercon,LisiNature2021}.

Here, we explore an alternative approach to Dirac / flat-band coupling based on proximity engineering in half-van-der-Waals (half-vdW) heterostructures. Specifically, we investigate graphene resting on a second two-dimensional (2D) material with localized electrons. This can be achieved by confinement heteroepitaxy, where intercalation of epitaxial graphene on SiC(0001) with Sn or Si creates 2D Mott-insulating flat-band systems beneath the graphene layer \cite{starke1998novel, johansson2000electronic, visikovskiy2016graphene, tilgner2025si, ghosal2025mott}. Alternatively, thermal decomposition of SiC$(0001)$ yields epitaxial graphene resting on a carbon buffer layer that also exhibits flat electronic bands \cite{berger2004ultrathin,riedl2010structural, emtsev2008interaction}. To date, experimental evidence of hybridization between the Dirac cone and the proximity-coupled flat bands remains elusive. 

We set out to provide such evidence using time- and angle-resolved photoemission spectroscopy (trARPES). We use femtosecond infrared and visible pulses to drive the heterostructures out of equilibrium and monitor their ultrafast relaxation dynamics directly in the band structure. The dynamics are compared to those of quasi-freestanding H-intercalated graphene on SiC$(0001)$, serving as a reference system without flat-band coupling \cite{riedl2009quasi, speck2011quasi, forti2011large}. We find that (1) carrier relaxation in all graphene / flat-band systems studied in the present work is approximately four times faster than in quasi-freestanding graphene, and that (2) graphene proximity-coupled to a flat-band system shows a transient charging with additional electrons, with decay times that systematically decrease with the atomic number of the atoms at the graphene-SiC interface. These observations point towards efficient interlayer interactions and tunable interlayer hybridization, consistent with recent theoretical predictions \cite{witt2026real-space}. Our results establish half-vdW heterostructures created by confinement heteroepitaxy as a versatile platform for the engineering of tunable Dirac / flat-band coupling and the exploration of exotic correlated phenomena such as heavy-fermion-like states. 

\section{Methods}
\subsection{sample preparation}

To grow graphene resting on a carbon buffer layer (G/C-SiC), n-doped, single-side polished 6H-SiC(0001) wafers (SiCrystal) were cleaned with acetone and isopropanol in an ultrasonic bath and subsequently treated with hydrofluoric acid. To remove the polishing defects, the substrates were H-etched at 1200\,$^{\circ}$C in a molecular hydrogen atmosphere for 6 minutes \cite{frewin2009}. Subsequent graphitization was carried out in an Aixtron high temperature BM cold-wall reactor in Ar atmosphere \cite{Emtsev2009}. Graphitization at 1200\,$^{\circ}$C resulted in the formation of a carbon-rich buffer layer characterized by a $(6\sqrt{3}\times6\sqrt{3})R30^\circ$ reconstruction. Increasing the temperature to 1270\,$^{\circ}$C for 5 to 10 minutes resulted in the formation of a graphene monolayer on top of the buffer layer (G/C-SiC). Alternatively, G/C-SiC was prepared using polymer-assisted sublimation growth (PASG) on epi-ready 4H-SiC substrates \cite{Kruskopf2016}. Prior to processing, the substrates were cleaned in acetone and isopropanol. A diluted polymer solution (AZ5214E photoresist/isopropanol = 0.0034) was deposited on the Si-face of the SiC substrates via spin-on deposition, enabling highly reproducible polymer coverage \cite{Chatterjee2022}. The samples were then placed in the growth chamber and annealed at 800\,$^{\circ}$C before growth. For buffer layer and graphene growth, the temperature was raised to 1400\,$^{\circ}$C and 1800\,$^{\circ}$C, respectively.

The growth of Sn-intercalated graphene (G/Sn-SiC), Si-intercalated graphene (G/Si-SiC) and quasi-freestanding H-intercalated graphene (G/H-SiC) started from a carbon buffer layer that was grown as described above \cite{Emtsev2009}. Sn was deposited from a Mo crucible with the substrate held at room temperature. Sn-intercalation was achieved by annealing the sample at 750\,$^{\circ}$C for 30 minutes, yielding a densely packed Sn $(1\times1)$ phase commensurate with the SiC $(1\times1)$ lattice. Further annealing at 1050\,$^{\circ}$C for 2 minutes resulted in the formation of the desired $(\sqrt{3}\times\sqrt{3})R30^\circ$ phase. Si was intercalated by annealing the sample at 750\,$^{\circ}$C while simultaneously exposing it to a Si flux from an EFM3 rod evaporator (Focus) for 4 hours. The Si flux was tracked with the attached flux monitor and kept constant at 0.1\,nA. H-intercalation was performed in a dedicated, contactless infrared heating system. The sample was annealed at 550\,$^{\circ}$C in an ultra-pure H$_2$ atmosphere (860\,mbar, 0.9\,slm) for 90 minutes.

\subsection{SPA-LEED}
A high-resolution spot-profile analysis low-energy electron diffraction (SPA-LEED) setup \cite{Scheithauer1986} was used to characterize the interface structures. All measurements were performed at room temperature using a primary electron energy of 150\,eV. The G/Sn-SiC sample was measured in situ. The G/Si-SiC sample was transferred to the SPA-LEED apparatus using a vacuum suitcase. The G/H-SiC sample was transferred in ambient conditions and subsequently cleaned by annealing at 400\,°C after reinsertion into ultrahigh vacuum.

\subsection{ARPES}
Static ARPES measurements were performed at room temperature with He I radiation and a hemispherical analyzer (Phoibos 100, Specs) that collected photoelectrons from a surface area of $\sim1$\,mm$^2$. Because the He I radiation was unpolarized, both branches of the Dirac cone were visible along the $\mathit{\Gamma K}$-direction of graphene, enabling the determination of the doping levels of the various graphene layers. The Gr/Si-SiC sample was transferred from the growth chamber into the ARPES chamber in a vacuum suitcase. All other samples were transferred in air and cleaned via annealing at $\sim600\,^{\circ}$C for $\sim5$\,minutes after reinsertion into ultra-high vacuum (UHV).

\subsection{trARPES}
Ultrafast laser pulses were generated by a Ti:Sapphire amplifier (Astrella, Coherent) delivering 35\,fs pulses at a central wavelength of 800\,nm with a repetition rate of 1\,kHz and a pulse energy of 7\,mJ. Two types of pump pulses were employed. In the first configuration, a fraction of the fundamental amplifier output was used directly as an infrared pump with photon energy $\hbar\omega_{pump}=1.55$\,eV. These pulses were used to excite G/Sn-SiC, G/Si-SiC and G/H-SiC with pump fluences of 0.5\,mJ/cm$^2$, 1.1\,mJ/cm$^2$ and 0.7\,mJ/cm$^2$, respectively. In the second configuration, pump pulses with $\hbar\omega_{pump}=2.0$\,eV were generated via frequency doubling of the output of an optical parametric amplifier (Topas Twins, Light Conversion). These pulses were used to excite G/C-SiC with a fluence of 1.1\,mJ/cm$^2$. The employed pump fluences represent a compromise between the goal to maximize the signal-to-noise ratio at the limited repetition rate of 1\,kHz and to minimize pump-induced space charge effects in the trARPES data. Probe pulses in the extreme ultraviolet (XUV) range were produced via high harmonic generation (HHG) in an argon gas jet, driven by the second harmonic of the amplifier output. A grating monochromator was used to isolate the 7th harmonic, corresponding to a probe photon energy of $\hbar\omega_{probe}=21.7$\,eV. The selected XUV beam was focused onto the sample using a toroidal mirror, yielding a spot diameter of approximately 400\,$\mu$m. trARPES snapshots were collected using the same hemispherical analyser as for the static ARPES measurements. Because the probe pulses were p-polarized, only one branch of the Dirac cone was visible in the trARPES data taken along the $\mathit{\Gamma K}$-direction of graphene \cite{ShirleyPRB1995, GierzPRB2011}. The energy resolutions were $230$\,meV (G/Sn-SiC), $220$\,meV (G/Si-SiC), $160$\,meV (G/C-SiC) and $200$\,meV (G/H-SiC). The temporal resolutions were $270$\,fs (G/Sn-SiC), $220$\,fs (G/Si-SiC), $100$\,fs (G/C-SiC) and $200$\,fs (G/H-SiC). 
	
\section{Results}

\begin{figure}
		\includegraphics[width = 1\columnwidth]{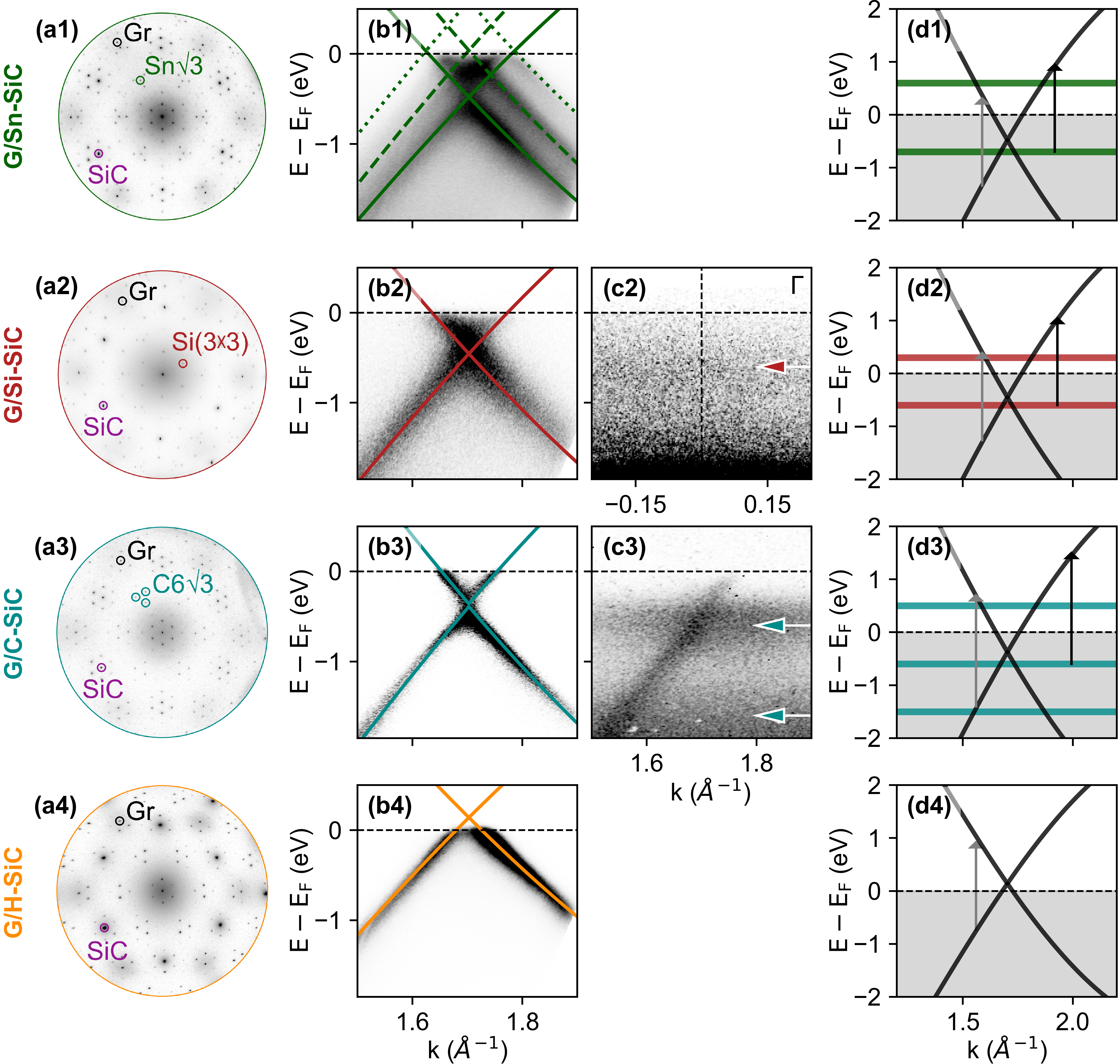}
		\caption{\textbf{Sample characterization.} \textbf{(1)} G/Sn-SiC. \textbf{(2)} G/Si-SiC. \textbf{(3)} G/C-SiC. \textbf{(4)} G/H-SiC. \textbf{(a)} SPA-LEED images taken at a kinetic energy of $150$\,eV. Labelled circles indicate the origin of various diffraction spots. \textbf{(b)} ARPES spectrum at $K$ along $\mathit{\Gamma K}$ direction using He I radiation. Solid lines are graphene tight-binding fits. In \textbf{(b1)} the dashed line indicates the dispersion of graphene on metallic $(1\times1)$-Sn. The dotted line indicates the dispersion of graphene resting on an unknown third phase. \textbf{(c2)} ARPES spectrum of G/Si-SiC at $\mathit{\Gamma}$ along $\mathit{\Gamma K}$ direction using He I radiation. \textbf{(c3)} ARPES spectrum at $K$ along $\mathit{\Gamma K}$ direction using XUV pulses at $21.7$\,eV of G/C-SiC with graphene coverage below one monolayer to provide access to the flat bands of the carbon}
		\label{figure1}
	\end{figure}

The results of the sample characterization are presented in Fig.\,\ref{figure1}. Fig.\,\ref{figure1}a shows SPA-LEED images of the three different half-vdW heterostructures and quasi-freestanding graphene. The diffraction spots of graphene and SiC(0001) are marked by black and pink circles, respectively. The diffraction spots of the various 2D flat-band systems are indicated in green (Sn-SiC), red (Si-SiC) and blue (C-SiC). All intercalates as well as the carbon buffer layer exhibit the desired structures characteristic of the various flat-band phases reported in literature: the $(\sqrt{3}\times\sqrt{3})R30^{\circ}$ and the $(3\times3)$ superstructure belonging to 2D Mott-insulating Sn \cite{mamiyev2022sn,mamiyev2024exploring,ghosal2025mott} and Si on SiC(0001) \cite{tilgner2025si}, respectively, and the $(6\sqrt{3}\times6\sqrt{3})R30^{\circ}$ reconstruction of the carbon buffer layer \cite{riedl2009quasi}. In the case of quasi-freestanding graphene (G/H-SiC in Fig.\,\ref{figure1}a4), the dangling bonds of the SiC substrate are saturated with H atoms, yielding a $(1\times1)$ surface structure \cite{riedl2009quasi, riedl2010structural}. 

The ARPES spectra in Fig.\,\ref{figure1}b were taken along the $\mathit{\Gamma K}$-direction at the $K$ point of graphene. The ARPES spectrum for Sn-intercalated graphene is shown in Fig.\,\ref{figure1}b1. It shows three Dirac cones with different doping levels, each originating from a distinct domain within the $\sim1$\,mm$^2$ field of view of the ARPES analyser. The Dirac cone with a doping level of $E_D\approx-480$\,meV highlighted by solid green lines is attributed to graphene resting on top of the Mott-insulating $(\sqrt{3}\times\sqrt{3})R30{^\circ}$ Sn-phase reported in \cite{mamiyev2022sn, ghosal2025mott}. The slightly p-doped Dirac cone (dashed green lines) is assigned to graphene resting on top of 2D metallic Sn with a $(1\times1)$ structure \cite{mamiyev2024exploring, kim2016charge,hayashi2017triangular, federl2025nonequilibrium}. The origin of the third Dirac cone marked by dotted green lines remains unclear. Note that Refs. \cite{VisikovskiyArxiv2018,LiPhysChemChemPhys2019} predict the existence of a third stable Sn-phase on SiC(0001) with a honeycomb lattice. All other half-vdW heterostructures exhibit a single Dirac cone, indicating the presence of a single structural domain within the field of view of the analyzer. The Dirac points are observed at $E_D\approx-450$\,meV (G/Si-SiC in Fig.\,\ref{figure1}b2), $E_D\approx-380$\,meV (G/C-SiC in Fig.\,\ref{figure1}b3) and $E_D\approx+140$\,meV (G/H-SiC in Fig.\,\ref{figure1}b4) in good agreement with literature \cite{xia2012si, visikovskiy2016graphene, tilgner2025si, emtsev2008interaction, riedl2010structural, forti2011large, riedl2009quasi}.

In Fig. \ref{figure1}c we present ARPES spectra of the flat bands originating from the 2D materials at the graphene-SiC interface. Note that the expected flat band of G/Sn-SiC at $E-E_F \approx - 700$\,meV \cite{ghosal2025mott} could not be detected, likely due to a poor cross section at the photon energy used. Further, G/H-SiC does not exhibit any flat bands as all dangling bonds are saturated with H atoms. The ARPES spectrum for G/Si-SiC at $\mathit{\Gamma}$ reveals a flat band at a binding energy of $E-E_F \approx - 600$\,meV in good agreement with literature \cite{johansson2000electronic, tilgner2025si}. To visualize the flat bands of G/C-SiC, we used a sample with a graphene coverage of less than one monolayer to increase the visibility of the band structure of the buffer layer. The ARPES spectrum in Fig. \ref{figure1}c3 clearly shows two flat bands at binding energies of $E-E_F\approx - 600$\,meV and $E-E_F\approx- 1.6$\,eV, in good agreement with literature \cite{emtsev2008interaction}.

The results of the ARPES measurements are summarized in Fig. \ref{figure1}d, where we sketch the Dirac cone and the flat bands for all samples. All flat-band systems also exhibit an unoccupied flat band the binding energy of which was obtained from literature \cite{ghosal2025mott, johansson2000electronic, GolerCarbon2013}. Note that the large linewidths of the bands prevent our ARPES measurements from resolving any avoided crossings between the Dirac cone and the occupied flat bands. Consequently, independent evidence for hybridization is required. 

To this end, we drive the half-vdW heterostructures out of equilibrium using femtosecond infrared or visible pump pulses and monitor the ensuing carrier dynamics directly in the band structure with trARPES. Under these conditions, we anticipate two distinct signatures of interlayer hybridization to emerge. First, photoexcitation excites carriers across the Dirac point (gray arrows in Fig.\,\ref{figure1}d). This generates a non-equilibrium carrier distribution inside the Dirac cone that is known to thermalize into a hot Fermi-Dirac distribution within tens of femtoseconds \cite{BreusingUltrafast2009, johannsen2015tunable}. The hot Fermi-Dirac distribution then cools down via the emission of optical and acoustic phonons on the picosecond timescale \cite{KampfrathPRL2005, YanPRB2009, johannsen2013direct}. The presence of a proximity-coupled 2D material is expected to accelerate the relaxation of the Dirac carriers by providing additional electronic and phononic relaxation channels \cite{federl2025nonequilibrium, Weigl_2025}. Second, interlayer hybridization turns the spectral weight of optical interlayer transitions non-zero, enabling the pump pulses to directly excite carriers from the flat bands into the Dirac cone (black arrows in Fig.\,\ref{figure1}d). This will result in a transient doping of the graphene layer. Subsequent carrier relaxation back to equilibrium then necessarily involves charge transfer that typically proceeds via hybridization-induced avoided crossings in the band structure, where the electronic wave function is delocalized over both layers \cite{LiChemistryOfMaterials2017, ZhuJMaterChemA2022, hofmann2023link}. The charge transfer rate is directly linked to the strength of the hybridization \cite{LiChemistryOfMaterials2017, ZhuJMaterChemA2022, hofmann2023link}.

\begin{figure}
		\includegraphics[width = 1\columnwidth]{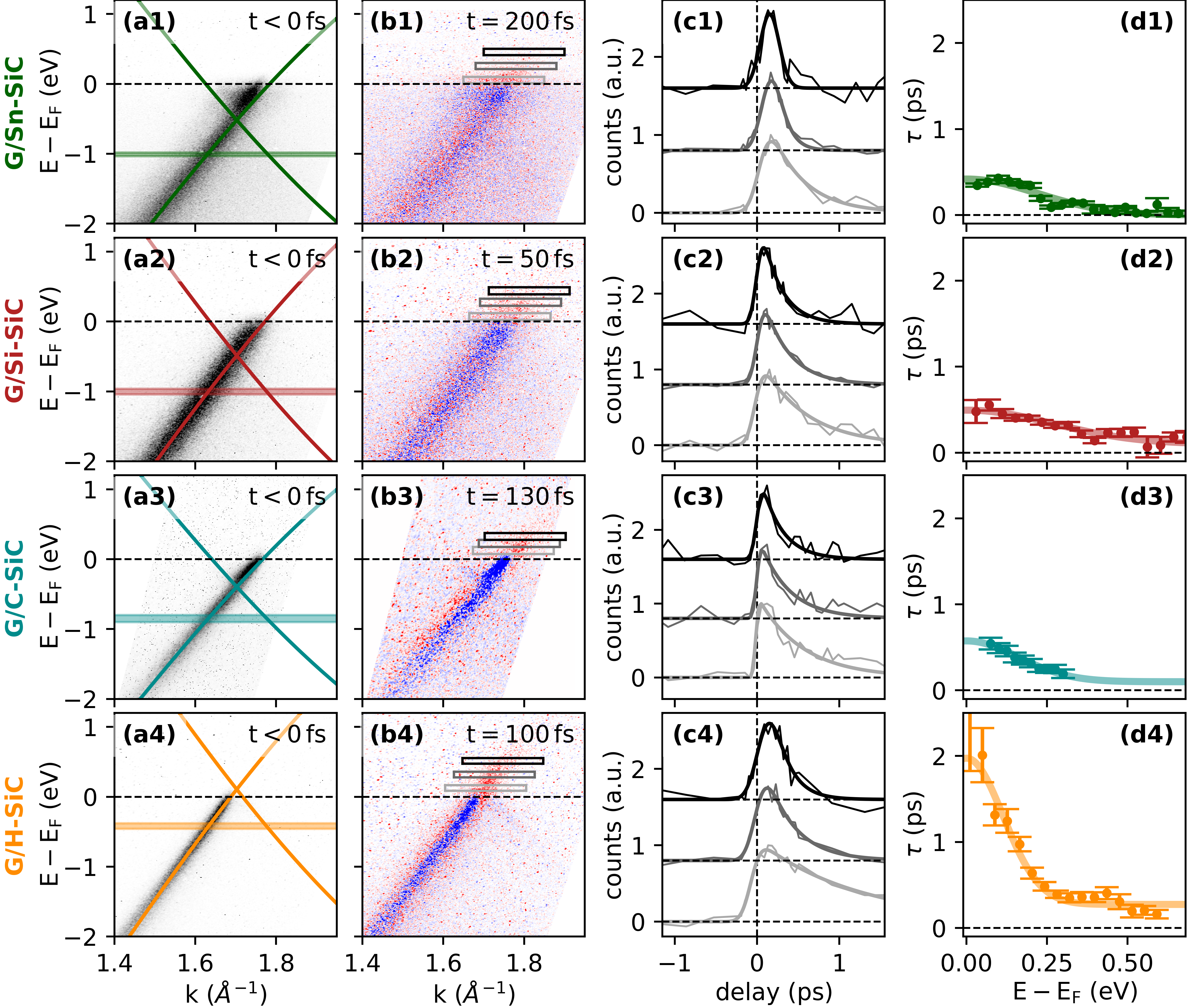}
		\caption{\textbf{Non-equilibrium carrier dynamics inside Dirac cone.} \textbf{(1)} G/Sn-SiC. \textbf{(2)} G/Si-SiC. \textbf{(3)} G/C-SiC. \textbf{(4)} G/H-SiC. \textbf{(a)} ARPES snapshot for negative pump-probe delay at $K$ along $\mathit{\Gamma K}$ direction of graphene using XUV pulses at $21.7$\,eV photon energy. Continuous lines indicate the tight binding dispersion of the Dirac cone. Shaded areas indicate the energies where the MDCs in Fig.\,\ref{figure3}a were extracted. \textbf{(b)} Pump-induced changes of the photocurrent at the peak of the pump-probe signal. Gain (loss) of photoelectrons is shown in red (blue). Boxes indicate the areas of integration for the pump-probe traces in (c). \textbf{(c)} Pump-probe traces obtained by integrating the counts over the areas marked by the boxes in (b) together with single-exponential fits. \textbf{(d)} Exponential lifetime $\tau$ from (c) as a function of energy. Continuous lines are Gaussian fits that serve as guides to the eye.}
	\label{figure2}
	\end{figure}

To explore the first scenario, we determine the energy-resolved population dynamics of the Dirac cone in Fig.\,\ref{figure2}. Figure\,\ref{figure2}a shows trARPES snapshots at $t<0$\,fs along the $\mathit{\Gamma K}$ direction at the $K$ point of graphene. The coloured lines are tight-binding fits confirming the doping levels of G/Sn-SiC ($E_D\approx-520$\,meV), G/Si-SiC ($E_D\approx-480$\,meV), G/C-SiC ($E_D\approx-400$\,meV) and G/H-SiC ($E_D\approx+100$\,meV). Note the presence of a single n-doped Dirac cone for G/Sn-SiC in Fig.\,\ref{figure2}a1, indicating that the spot size of the XUV probe pulses is small enough to address a single $\sqrt{3}$ domain on the sample surface. Fig.\,\ref{figure2}b shows the pump-induced changes of the photocurrent at the peak of the pump-probe signal. All samples show a gain of photoelectrons (red) above the Fermi level and a loss of photoelectrons (blue) below the Fermi level with respect to negative pump-probe delay times. To gain access to the energy-resolved lifetimes we integrate the photocurrent over the areas marked by the boxes in Fig.\,\ref{figure2}b. Figure\,\ref{figure2}c shows the resulting pump-probe traces together with single-exponential fits (see Supplemental Material). Similar fits were performed for other energies above the Fermi level. Fig.\,\ref{figure2}d shows the resulting energy-dependent relaxation times $\tau$ together with Gaussian fits that serve as guides to the eye (solid lines). We find that, for all samples, the lifetimes increase as we approach the Fermi level as expected for a cooling Fermi-Dirac distribution \cite{GierzFaradayDiscuss2014, gierz2015population, johannsen2015tunable}. The peak relaxation time, $\tau_{max}$, is found to be $\sim0.5$\,ps for all graphene layers resting on a flat-band system (Fig.\,\ref{figure2}d1-d3) and $\sim2$\,ps for G/H-SiC (Fig.\,\ref{figure2}d4). This fourfold decrease of the relaxation times compared to quasi-freestanding G/H-SiC provides first evidence of interlayer interactions between graphene and the 2D flat-band systems.

\begin{figure}
		\includegraphics[width =0.84\columnwidth]{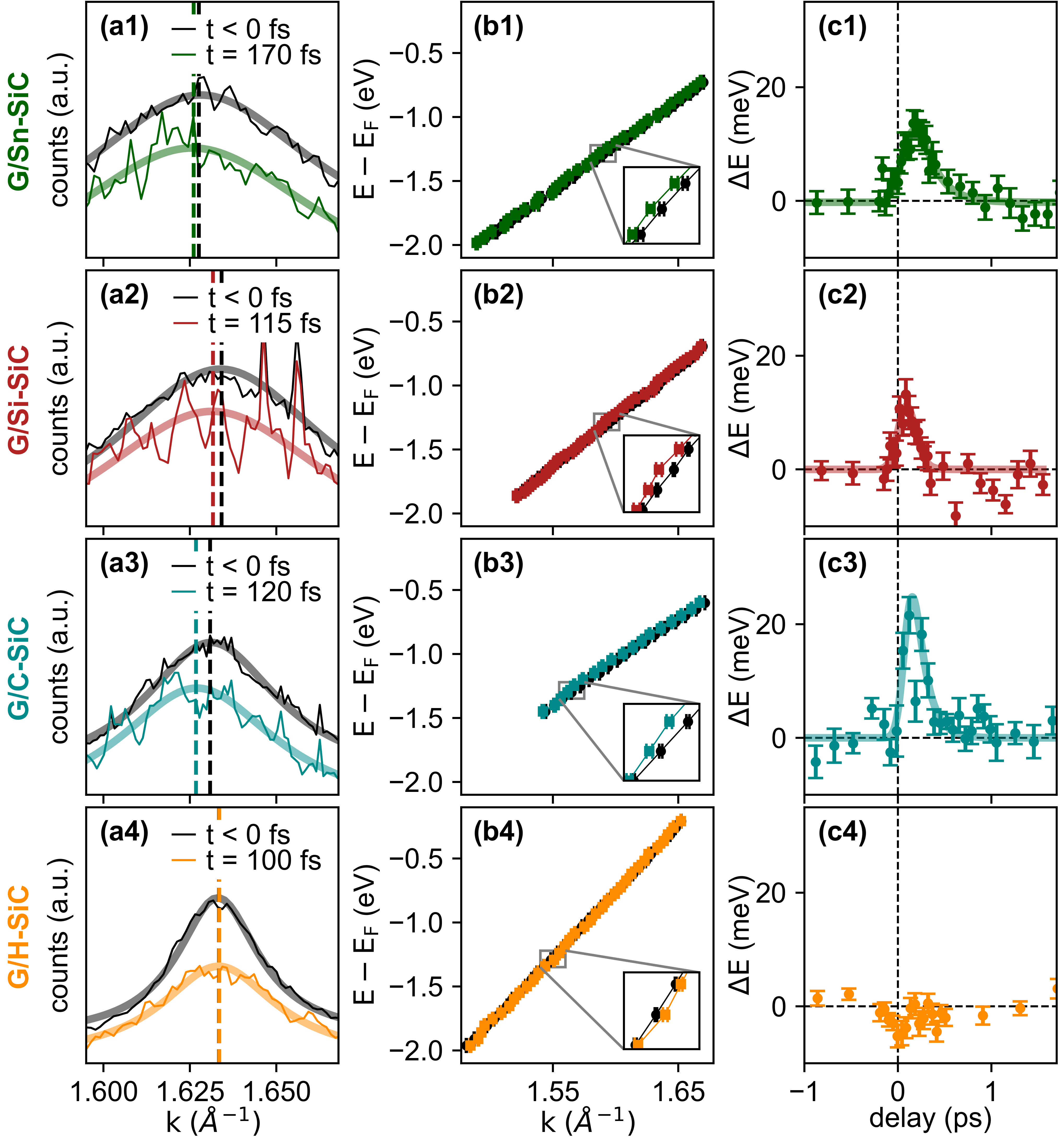}
		\caption{\textbf{Transient band structure} \textbf{(1)} G/Sn-SiC. \textbf{(2)} G/Si-SiC. \textbf{(3)} G/C-SiC. \textbf{(4)} G/H-SiC. \textbf{(a)} MDCs extracted at the energies indicated in Fig.\,\ref{figure2}a at negative pump-probe delay (black) and at the peak of the pump-probe signal (green, red, blue and orange) with Lorentzian fits. \textbf{(b)} Resulting dispersion of the Dirac cone at negative pump-probe delay (black) and at the peak of the pump-probe signal (green, red, blue and orange). \textbf{(c)} Transient shift of Dirac cone, $\Delta E$, as a function of pump-probe delay together with single-exponential fit that serves as guide to the eye.}
	\label{figure3}
	\end{figure}

To address the second scenario, we first search for transient changes in the binding energy of the Dirac cone caused by a possible charging of the graphene layer with additional electrons. To this end, we extract momentum distribution curves (MDCs) at various energies across the Dirac cone that we fit with a Lorentzian (see Fig.\,\ref{figure3}a). This yields the dispersion of the Dirac cone displayed in Fig.\,\ref{figure3}b that we fit with a straight line, the slope of which is fixed to the value obtained for negative pump-probe delays (see Supplemental Material). Figure\,\ref{figure3}c displays the shift of the Dirac cone, $\Delta E (t)$, as a function of pump-probe delay together with a single-exponential fit (see Supplemental Material) that serves as guide to the eye. We find that all graphene / flat-band systems exhibit a transient upshift of the Dirac cone with an amplitude of $\sim$10\,meV (rows 1-3 in Fig.\,\ref{figure3}), while no shift is observed in the case of quasi-freestanding G/H-SiC (row 4 in Fig.\,\ref{figure3}). The observed transient reduction in binding energy $E-E_F$ is consistent with a transient negative charging of the graphene layer.
 
To determine the transient carrier concentration inside the various Dirac cones, we integrate the trARPES snapshots from Fig.\,\ref{figure2}a over the momentum axis, yielding the energy distribution curves (EDCs) shown in Fig.\,\ref{figure4}a. In the vicinity of the Fermi level, the EDCs are fitted with a Fermi-Dirac distribution (see Supplemental Material). The resulting shift of the chemical potential with respect to the Dirac point, $\Delta\mu_{e}$, and the electronic temperature, $T_e$, are shown in Fig.\,\ref{figure4}b. From the data in Fig.\,\ref{figure4}b we then calculate the transient carrier concentration inside the Dirac cone $\Delta n_e= \int_{-\infty}^{\infty}\rho(E)[f_{FD}(E, \mu_e(t), T_e(t))-f_{FD}(E,\mu_e(t<0), T_e(t<0))]dE $ for every pump-probe delay, where $\rho(E)$ is the density of states of graphene. The results are shown in Fig.\,\ref{figure4}c, together with single-exponential fits (see Supplemental Material). We find that all graphene layers resting on a flat-band system become negatively charged with an increase in carrier concentration on the order of $10^{12}/$cm$^2$ (Figs.\,\ref{figure4}c1-c3). Furthermore, we observe that the lifetime of the transient carrier concentration decreases from $\tau=440\pm100$\,fs in G/Sn-SiC, to $\tau=310\pm50$\,fs in G/Si-SiC to $\tau=100\pm40$\,fs in G/C-SiC, i.e. with decreasing atomic number of the atomic species at the graphene-SiC interface. The charge carrier concentration in quasi-freestanding G/H-SiC, on the other hand, is found to remain constant throughout the excitation and relaxation process (Fig.\,\ref{figure4}c4). Consistent with the second scenario and the band shifts observed in Fig.\,\ref{figure3}, we find that all graphene layers resting on a flat-band system experience a transient charging with additional electrons, the decay times of which shorten with decreasing atomic number of the atoms at the graphene-SiC interface.

According to literature \cite{LiChemistryOfMaterials2017, ZhuJMaterChemA2022, hofmann2023link}, the timescale for ultrafast charge transfer responsible for the decay of the transient doping of the Dirac cone is determined by the size of the hybridization-induced avoided crossings and the available phase space. As the band alignment (and therefore the scattering phase space) across all three investigated graphene / flat-band systems is similar (Fig.\,\ref{figure1}d), we attribute the observed decrease of the relaxation time with decreasing size of the atoms at the graphene-SiC interface to an increase of the hybridization strength. This interpretation is supported by recent theoretical work \cite{witt2026real-space} where the authors use density functional theory to investigate the electronic structure of graphene on top of different $(\sqrt{3}\times\sqrt{3})R30^{\circ}$ adsorbate structures on SiC(0001) hosting flat bands. They show that, the smaller the size of the adsorbate, the closer it will be to the graphene layer, resulting in a hybridization strength that increases with decreasing size of the adsorbate.

Finally, we would like to point out that all graphene layers resting on a flat-band system are n-doped, whereas quasi-freestanding G/H-SiC is p-doped. Some of us performed a systematic study of the doping dependence of the non-equilibrium carrier dynamics in graphene in the past \cite{Weigl_2025} which allows us to rule out differences in doping level as a possible explanation for the distinct dynamics observed in the present work.

\section{Summary}

In summary, using trARPES we have provided direct evidence for the presence of hybridization between the graphene Dirac cone and the flat bands of various proximity-coupled 2D surface structures on SiC(0001). This hybridization manifests in three central observations: (i) a pronounced reduction in the carrier lifetime, reflecting additional electronic and phononic relaxation channels enabled by the flat-band subsystem; (ii) an increase in carrier density within the Dirac cone, corresponding to a hybridization-enabled optical excitation from the flat bands into graphene; and (iii) subsequent decharging dynamics revealing the strength of the interlayer coupling, which increases with decreasing atomic number of the intercalated atoms at the graphene-SiC interface. Together, these findings establish this materials platform as a promising route toward realizing exotic heavy-fermion behavior.

\section{Acknowledgments}

This work received funding from the European Union’s Horizon 2020 research and innovation program under Grant Agreement No. 851280-ERC-2019-STG as well as from the Deutsche Forschungsgemeinschaft (DFG) via the collaborative research centre CRC 1277 (Project No. 314695032), and the Research Unit RU 5242 (Project No. 449119662).

\clearpage

\pagebreak

\bibliography{literature}% Produces the bibliography via BibTeX.

\pagebreak
	
\clearpage
\pagebreak

\noindent buffer layer. Coloured arrows in (c2) and (c3) indicate the binding energies of the flat bands. \textbf{(d)} Sketch of band structure showing Dirac cone of graphene and flat bands of 2D material at graphene-SiC interface. Graphene interband transitions and transitions from flat bands to graphene Dirac cone are indicated by grey and black arrows, respectively.

\clearpage
\pagebreak

	\begin{figure}
		\includegraphics[width = 0.88\columnwidth]{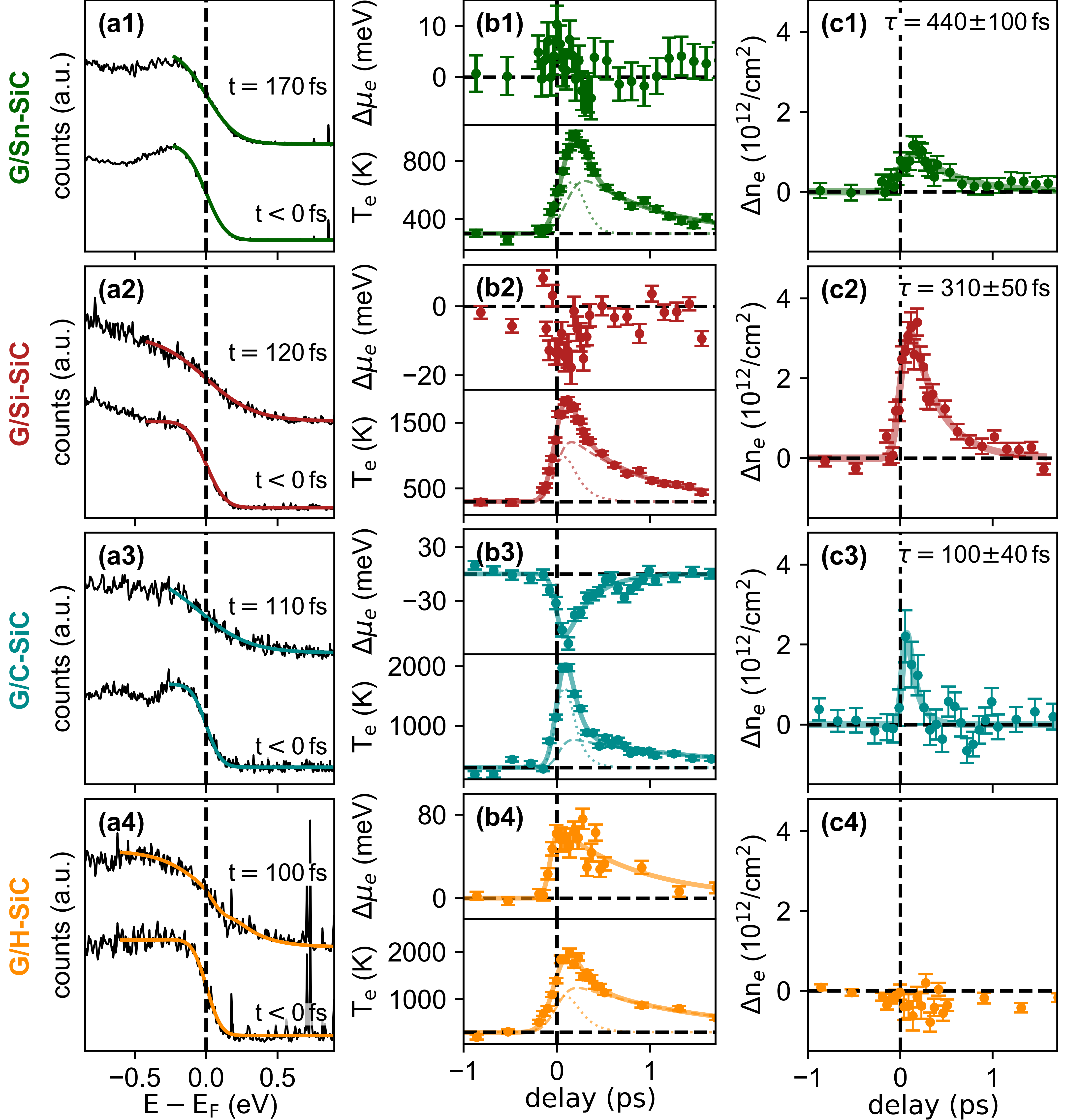}
		\caption{\textbf{Ultrafast charge transfer.} \textbf{(1)} G/Sn-SiC. \textbf{(2)} G/Si-SiC. \textbf{(3)} G/C-SiC. \textbf{(4)} G/H-SiC. \textbf{(a)} Energy distribution curves obtained by integrating the photoemission current over the momentum axis together with Fermi-Dirac fits, as described in the main text. \textbf{(b)} \textbf{top:} Resulting transient chemical potential shift together with single-exponential fit. \textbf{bottom:} Transient electronic temperature together with double-exponential fit (see Supplemental Material). The fast and slow decay components commonly attributed to the emission of optical and acoustic phonons, respectively, are indicated by dotted and dashed lines. \textbf{(c)} Transient change of the carrier concentration inside the Dirac cone together with single-exponential fit (c1-3).}
	\label{figure4}
	\end{figure}
	
\clearpage
\pagebreak

\section{Supplementary Material}

\subsection{Fit functions for single- and double-exponential decay}
	
To extract rise and decay times of a given pump-probe signal, the following analytic fitting function is used that stems from convolving the product of a step function and an exponential decay for the underlying dynamics with a Gaussian function accounting for the finite temporal resolution:
	
	\begin{equation}
		f(t) = \frac{a}{2} \left( 1 + \mathrm{erf}\left( \frac{(t-t_0)\tau - \frac{\text{FWHM}^2}{8\ln2}}{\sqrt{2}\tau\frac{\text{FWHM}}{2\sqrt{2\ln2}}} \right) \right)
		\exp\left( \frac{\frac{\text{FWHM}^2}{8\ln 2}-2(t-t_0)\tau}{2\tau^2} \right)
	\end{equation}
	
\noindent Here, $a$ is the amplitude, $\text{FWHM}$ is the full width at half maximum of the Gaussian, $t_0$ is the time delay at which pump and probe pulses overlap and $\tau$ is the decay time.\\
	
\noindent For pump-probe signals that contain two exponentially decaying contributions with different decay times, the following double-exponential decay function is used:
	
	\begin{equation}\begin{split}
			f(t) = \frac{A}{2} \left[a \left(1 + \mathrm{erf}\left(
			\frac{(t - t_0) \tau_1 - \frac{\text{FWHM}^2}{8\ln2}}{\sqrt{2}\tau_1 \frac{\text{FWHM}}{2\sqrt{2\ln2}}}\right)\right) \exp\left( \frac{\frac{\text{FWHM}^2}{8\ln 2}-2(t-t_0)\tau_1}{2\tau_1^2} \right) \right.+\\
			+\left. (1 - a) \left(1 + \mathrm{erf}\left(
			\frac{(t - t_0) \tau_2 - \frac{\text{FWHM}^2}{8\ln2}}{\sqrt{2}\tau_2 \frac{\text{FWHM}}{2\sqrt{2\ln2}}}\right)\right) \exp\left( \frac{\frac{\text{FWHM}^2}{8\ln 2}-2(t-t_0)\tau_2}{2\tau_2^2} \right)\right]
	\end{split}\end{equation}
	
\noindent Here, $A$ is the main amplitude, $0\leq a\leq1$ is the relative weight of the first decay with respect to the second decay and $\tau_1$ and $\tau_2$ are the decay times.

\subsection{Extracting transient shift of Dirac cone}

\begin{figure}[h]
		\includegraphics[width = 1\columnwidth]{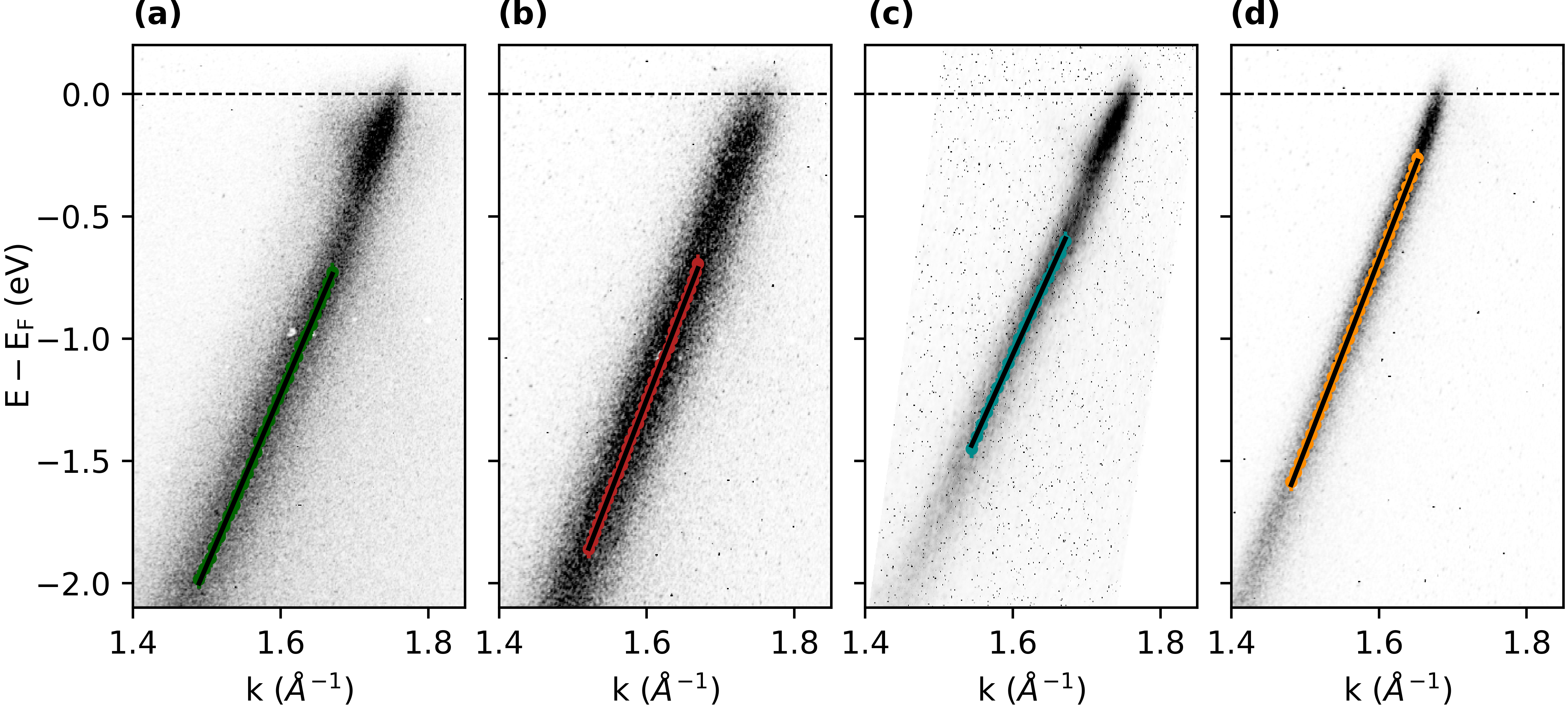}
		\caption{\textbf{Dirac cone dispersion} at negative pump-probe delay. \textbf{(a)} G/Sn-SiC. \textbf{(b)} G/Si-SiC. \textbf{(c)} G/C-SiC. \textbf{(d)} G/H-SiC. Coloured data points indicate peak positions obtained from MDC fits. Black lines are linear fits of the MDC peak positions.}
		\label{figure_S1}
	\end{figure}

The linear fits used to extract the transient shift of the Dirac cone in the different graphene samples in Fig.\,3b of the manuscript are shown in Fig.\,\ref{figure_S1}.

\subsection{Fermi-Dirac fits}
ARPES snapshots were taken along the $\mathit{\Gamma K}$ direction of graphene. For an infinitesimally thin cut through the K point, the density of states (DOS) contributing to the ARPES snapshot is independent of energy. The finite angular resolution, however, leads to an integration over a small range of $k$-values perpendicular to the nominal $\mathit{\Gamma K}$ direction. This introduces an energy dependence into the measured DOS (see Fig.\,\ref{figure_S1}a) that needs to be considered when performing Fermi-Dirac fits of the transient population of the Dirac cone. This effect is particularly sever for slightly p-doped G/H-SiC where the measured Dirac cone population, extracted by integrating the counts over the full momentum range of Fig.\,\ref{figure_S1}b, shows as clear dip at the Dirac point (see Fig.\,\ref{figure_S1}c). In this case, the fitting function used to extract the transient electronic temperature and chemical potential was the following:

$$f(E)=(DOS(E)\cdot f_{FD}(E,\mu_e,T_e))\ast Gaussian,$$

\noindent where $DOS(E)$ is the density of states sketched in red in Fig.\,\ref{figure_S1}a, $f_{FD}(E,\mu_e,T_e)$ is the Fermi-Dirac distribution, and the convolution with the Gaussian accounts for the finite energy resolution of the experiment.

All other samples explored in the manuscript are n-doped with the Dirac point several 100\,meV below $E_F$. In this case, the DOS that contributes to the ARPES spectra is approximately constant in the vicinity of the Fermi level and the fitting function simplifies to 

$$f(E)=f_{FD}(E,\mu_e,T_e)\ast Gaussian,$$

\noindent where the amplitude of the Fermi-Dirac distribution was fixed to the value obtained for negative pump-probe delays.

To reference the chemical potential to the Dirac point, we subtracted the transient band shift $\Delta E(t)$ from Fig.\,3c in the manuscript from the fit result for the chemical potential $\mu_e$.
	
		\begin{figure}[h]
		\includegraphics[width = 1\columnwidth]{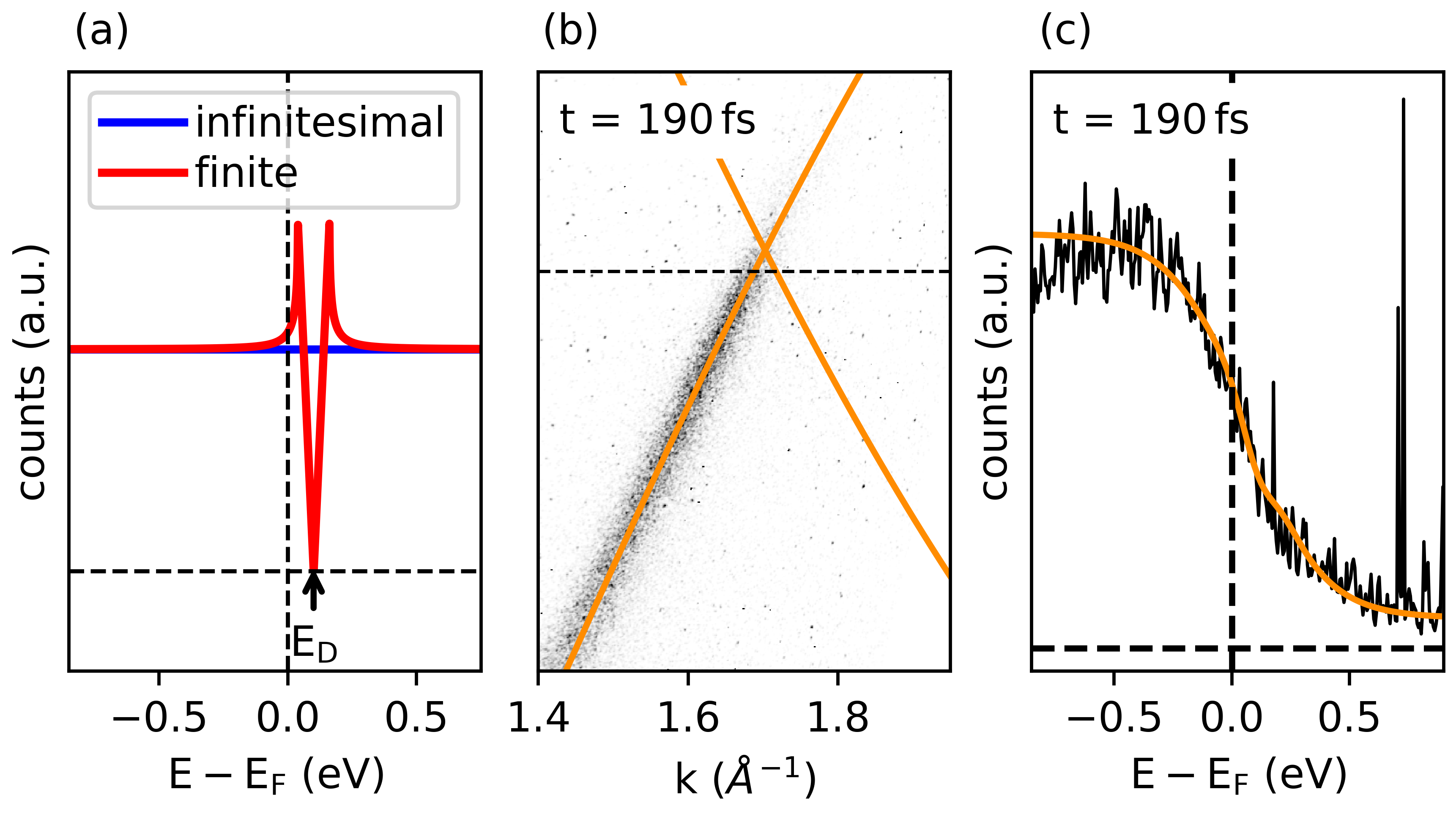}
		\caption{\textbf{Fermi-Dirac fits for G/H-SiC.} \textbf{(a)} Density of states as a function of energy for an infinitesimally thin cut along the $\mathit{\Gamma K}$ direction of graphene (blue) and for a cut with finite width (red). \textbf{(b)} ARPES snapshot of G/H-SiC at a pump-probe delay of $t=190$\,fs. \textbf{(c)} Population of the Dirac cone as a function of energy obtained by integrating the ARPES snapshot from (b) over the momentum axis together with Fermi-Dirac fit including the energy-dependent DOS from (a).}
		\label{figure_S2}
	\end{figure}

	\end{document}